\documentstyle[prl,preprint,aps,epsfig]{revtex}
\tightenlines
\begin{document}

\newcommand{\adag}{a^{\dag}}
\newcommand{\atil}{\tilde{a}}
\def\frp#1{${#1\over2}^+$}
\def\frm#1{${#1\over2}^-$}
\def\g{\noindent}

\def\mev{\hbox{\ MeV}}
\def\kev{\hbox{\ keV}}
\def\lambdabar{{\mathchar'26\mkern-9mu\lambda}}
\def\lambdabarrr{{^-\mkern-12mu\lambda}}
 
\draft
\title{Shell model description of $^{16}$O(p,$\gamma$)$^{17}$F and 
$^{16}$O(p,p)$^{16}$O reactions}

\author{K. Bennaceur$^{1,2}$, N. Michel$^{1}$, F. Nowacki$^{3}$, 
J. Oko{\l}owicz$^{1,4}$ ~and M. P{\l}oszajczak$^{1}$}
\address{1.\ Grand Acc\'{e}l\'{e}rateur National d'Ions Lourds (GANIL),
CEA/DSM -- CNRS/IN2P3, BP 5027, F-14076 Caen Cedex 05, France}
\address{2.\ Centre d'Etudes de Bruy\`{e}res-le-Ch\^{a}tel, 
BP 12, F-91680 Bruy\`{e}res-le-Ch\^{a}tel, France}
\address{3.\ Laboratoire de Physique Th\'{e}orique  Strasbourg (EP 106),
3-5 rue de l'Universite, F-67084 Strasbourg Cedex, France }
\address{4.\ Institute of Nuclear Physics, Radzikowskiego 152,
PL - 31342 Krakow, Poland}

%\date{\today}

\maketitle

\begin{abstract}
\parbox{14cm}{\rm We present shell model calculations of both the
structure of $^{17}$F and the reactions 
$^{16}$O(p,$\gamma$)$^{17}$F,  
$^{16}$O(p,p)$^{16}$O. We use the
ZBM interaction which provides a fair description of the properties of
$^{16}\mbox{O}$ and neighbouring nuclei and, in particular it takes account for
the complicated correlations in coexisting low-lying states of
$^{16}\mbox{O}$.}
\end{abstract}
\bigskip

\vfill
\newpage

A realistic account of the low-lying states properties 
in exotic nuclei requires taking into
account the coupling between discrete and continuum states which is responsible
for unusual spatial features of these nuclei. Within the newly
developed Shell Model Embedded in the Continuum (SMEC) approach \cite{bnop1}, 
one may obtain
the unified description of the divergent characteristics of these states as
well as
the reactions involving one-nucleon in the continuum. This provides a
stringent test of approximations involved in the SMEC calculations and
permits to asses the mutual complementarity of the reaction and structure data
for understanding of these nuclei. The quality of the SMEC
description depends crucially on the realistic account of the 
configuration mixing for coexisting low-lying structures and hence on the
quality of the Shell Model (SM) effective interactions and the SM space
considered. In this work, we shall present the calculation for $^{17}\mbox{F}$
for which it is believed that configurations of up to four particles and four
holes are necessary. Closed shell nuclei are never inert and multiple 
particle-hole excitations are always observed in their spectra.
In $^{16}$O and $^{17}$O, Brown and Green described the low-lying spectra 
by mixing spherical and deformed states \cite{brown}.
From the shell model point of view, Zuker-Buck-McGrory (ZBM) set 
an effective interaction in the basis 
of $0p_{1/2}$, $1s_{1/2}$ and $0d_{5/2}$ orbitals \cite{zbm,zbm1}. 
This valence space has the advantage
to be practically non spurious and most of states at the $p-sd$ interface 
around $^{16}$O are nicely described through configuration mixing of these 
three orbitals. In particular, the energy spectra, spectroscopic factors  
and correlations in the
low-lying states of $A=16$ and $A=17$ nuclei are well reproduced 
\cite{comment}. The
wavefunction components for the first three $0^+$ states in $^{16}$O are in a
fair agreement with the recently developed interactions in the full $p-sd$
shells \cite{smforce}. The aim of this work is not to provide better new SM
wavefunctions for $^{16}$O but to build on them the continuum effects and to
investigate the consequences of this coupling both for the structure of $^{17}$F
and for the reactions $^{16}$O(p,$\gamma$)$^{17}$F,  
$^{16}$O(p,p)$^{16}$O. For that purpose, the ZBM interaction is
satisfactory and we are going to use it in this study. 

In the SMEC formalism  
the subspaces of (quasi-) bound (the $Q$ subspace) and scattering (the
$P$ subspace) states are not separated artificially \cite{bnop1}. 
(For the review of earlier works see also
Ref. \cite{bartz3} ). Using
the projection operator technique, we separate the $P$ subspace of
asymptotic channels from the $Q$ subspace of many-body localized
states which are
build up by the bound single-particle (s.p.) wavefunctions
and by the s.p.\ resonance wavefunctions. $P$ subspace  
contains $(N-1)$-particle localized states 
and one nucleon in the scattering state. The s.p.\
resonance wavefunctions outside of the cutoff radius $R_{cut}$ are
included in the $P$ subspace. The resonance wavefunctions for
$r < R_{cut}$~ are included in the $Q$ subspace. The
wavefunctions in $Q$ and $P$ are then properly renormalized
in order to ensure the orthogonality of wavefunctions in both subspaces.

In the first
step, we calculate the (quasi-) bound many-body states in $Q$ subspace. For
that we solve the multiconfigurational SM problem :
$H_{QQ}{\Phi}_i = E_i{\Phi}_i$,
using the code ANTOINE \cite{caurier}. For $H_{QQ} \equiv QHQ$ we take
the ZBM interaction which yields realistic {\it internal mixing} of many-body
configurations in $Q$ subspace. 

To generate the radial s.p.\ wavefunctions in $Q$ subspace
and the scattering wavefunctions in $P$ subspace
we use the average potential of Woods-Saxon (WS) type
with the spin-orbit and Coulomb parts included:
$$U(r) = V_0f(r) + V_{SO} {\lambdabar}_{\pi}^2 (2{\bf l}\cdot{\bf s})
\frac{1}{r}\frac{d{f}(r)}{dr}  + V_C  ~,$$
where ${\lambdabar}_{\pi}^2 = 2\,$fm$^2$ is the pion Compton wavelength and
$f(r)$ is the spherically symmetrical WS formfactor :
$f(r) = \left[ 1 + \exp ((r-R_0)/a) \right]^{-1}$.
The Coulomb potential $V_C$ 
is calculated for the uniformly charged sphere with radius $R_0$. This 
'first guess' potential $U(r)$, is then modified by the
residual interaction. We shall return to this problem below.

For the continuum part, we solve the coupled channel equations :
$$(E^{(+)} - H_{PP}){\xi}_{E}^{c(+)} \equiv
\sum_{c^{'}}^{}(E^{(+)} - H_{cc^{'}}) {\xi}_E^{c^{'}(+)} = 0 ~,$$
where index $c$~ denotes different channels and $H_{PP} \equiv PHP$~.
The superscript $(+)$ means that boundary
conditions  for incoming wave in the channel $c$ and
outgoing scattering waves in all channels are used.
The channel states are defined by coupling of one
nucleon in the scattering continuum to the many-body SM state in
$(N - 1)$-nucleus. For the coupling between bound and scattering
states around $^{16}\mbox{O}$, we use 
the density dependent interaction  
which is close to the Landau - Migdal type of interactions 
\cite{schw84,ostatnia}. However, 
as compared to the original force of Schwesinger and Wambach 
\cite{schw84}, the radius parameter $r_0$ 
of the WS density formfactor  
is somewhat reduced to better fit the experimental 
matter radius in oxygen ($r_0=2.64$ fm). This interaction provides {\it
external mixing} of SM configurations via the virtual excitations of
particles to the continuum states.

The channel - channel coupling potential is :
\begin{eqnarray}
\label{esp2}
H_{cc^{'}} = (T + U ){\delta}_{cc^{'}} + {\upsilon }_{cc^{'}}^{J} ~ \ ,
\end{eqnarray}
where $T$ is the kinetic-energy operator and
$ {\upsilon }_{cc^{'}}^{J}$ is the channel-channel coupling generated by the
residual interaction. Reduced
matrix elements of the channel - channel coupling, 
which involve one-body operators of the kind :
${\cal O}^{K}_{{\beta}{\delta}} =
(a^{\dagger}_{\beta}{\tilde a}_{\delta})^K$, 
depend sensibly on the amount of $2p - 2h$ and
$4p - 4h$ correlations in the ground state of $^{16}\mbox{O}$.
The potential for channel $c$
in (\ref{esp2}) consists of initial WS guess, $U(r)$, and of the diagonal
part of coupling potential ${\upsilon }_{cc}^{J}$
which depends on both the s.p.\ orbit
${\phi}_{l,j}$ and the considered many-body
state $J^{\pi}$. This modification of the initial potential $U(r)$ change the
generated s.p.\ wavefunctions ${\phi}_{l,j}$ defining $Q$ subspace which in
turn modify the diagonal part of the residual force, {\it etc.}
In other words, the procedure of solving of the coupled channel equations is
accompanied by the self-consistent iterative procedure which yields for each
total $J$ independently the new self-consistent potential :
$$U^{(sc)}(r) = U(r)+{\upsilon }_{cc}^{J(sc)}(r) ~,$$
and consistent with it the new renormalized formfactor of the
coupling force. $U^{(sc)}(r)$ differs significantly 
from the initial WS potential,
especially in the interior of the potential \cite{bnop1,ostatnia}. 
Parameters of $U(r)$
are chosen  in such a way that $U^{(sc)}(r)$
reproduces energies of experimental s.p.\ states,
whenever their identification is possible.

 The third system of equations in SMEC consists of the inhomogeneous
coupled channel equations:
$$(E^{(+)} - H_{PP}){\omega}_{i}^{(+)} = H_{PQ}{\Phi}_i \equiv w_i$$
with the source term $w_i$ which depends on the
structure of $N$ - particle SM wavefunction ${\Phi}_i$. 
  Formfactor of the source term is given by
the self-consistently determined s.p.\ wavefunctions.
The solutions : ${\omega}_{i}^{(+)} \equiv G_P^{(+)}H_{PQ}\Phi_i$, where 
$G_{P}^{(+)}$~ is the Green function for the motion of s.p.\ in
the $P$ subspace, 
describe the decay of quasi-bound state ${\Phi}_i$~ in the continuum.
Reduced matrix elements of the source term, which involve products of two
annihilation operators and one creation operator of the kind :
${\cal R}^{j_{\alpha}}_{{\gamma}{\delta}(L){\beta}}=
(a^\dagger_{\beta}({\tilde a}_{\gamma}{\tilde a}_{\delta})^L)^{j_{\alpha}}$,
are calculated between different initial state wavefunctions 
in $^{17}\mbox{F}$ and a given final state wavefunction in $^{16}\mbox{O}$.
It should be stressed that the matrix elements of the source term depend
sensitively on the percentage of the shell closure in $^{16}$O, {\it i.e.}, on
the amount of correlations both in the g.s. of
$^{16}$O and in the considered states of $^{17}\mbox{F}$. Obviously, this kind
of couplings are not accounted for by the spectroscopic amplitudes.

The total wavefunction is expressed by three functions:
${\Phi}_i$~, ${\xi}_{E}^{c}$~ and ${\omega}_i$ \cite{bnop1,bartz3} :
\begin{eqnarray}
\label{eq2}
{\Psi}_{E}^{c} = {\xi}_{E}^{c} + \sum_{i,j}({\Phi}_i + {\omega}_i)
\frac{1}{E - H_{QQ}^{eff}}
<{\Phi}_{j}\mid H_{QP} \mid{\xi}_{E}^{c}> ~ 
\end{eqnarray}
where : $H_{QQ}^{eff}(E) = H_{QQ} + H_{QP}G_{P}^{(+)}H_{PQ} ~,$
is the new energy dependent effective
SM Hamiltonian which contains the coupling to the continuum.
Operator $H_{QQ}^{eff}(E)$, which is Hermitian for
energies below the particle emission threshold, becomes non-Hermitian 
for energies higher than the threshold. Consequently,
the eigenvalues ${\tilde {E_i}} - \frac{1}{2}i{\tilde {{\Gamma}_i} }$ are
complex for decaying states and
depend on the energy $E$ of the particle in the continuum.
The energy and the width of resonance states are determined by the condition:
$\tilde{E_i}(E) = E$. The eigenstates
corresponding to these eigenvalues can be obtained by the orthogonal but in
general non-unitary transformation \cite{bartz3,ostatnia}.
Inserting them in (\ref{eq2}), one obtains symmetrically 
the new continuum many-body 
wavefunctions modified by the discrete states, and the new
discrete state wavefunctions modified by the coupling 
to the continuum states.

The SMEC wavefunctions, can be used to
calculate various spectroscopic and reaction quantities. These include for example
 the proton (neutron) capture data, Coulomb dissociation data, 
elastic (inelastic) proton (neutron) scattering data,
energies and wavefunctions of discrete and resonance states, transition matrix
elements between SMEC  wavefunctions, static nuclear 
moments {\it etc.} \cite{bnop1,ostatnia,shyam}. 
The application of the SMEC model  for the
description of structure  for mirror nuclei
and capture cross sections for mirror reactions in $p$-shell
has been published in Ref. \cite{bnop1}. The analysis of the structure 
of $^{17}\mbox{F}$ and the reactions 
$^{16}\mbox{O}(\mbox{p},\gamma )^{17}\mbox{F}$,  
$^{16}\mbox{O}(\mbox{p,p})^{16}\mbox{O}$ in the $(0p1s0d)$-space, 
{\it neglecting} the 2p-2h, 4p-4h admixtures in$^{16}\mbox{O}$ wavefunctions 
have been reported recently as well \cite{ostatnia}. To correct for the missing
correlations in the low-energy wavefunctions of $^{16}$O and $^{17}$F, the matrix
elements of the $Q - P$ coupling in this study \cite{ostatnia} 
have been quenched by the
factors related to the spectroscopic amplitudes for positive parity states in
$^{17}\mbox{F}$ ($^{17}\mbox{O}$) and to the amount of $2p - 2h, 4p - 4h$
correlations in the g.s.\ of $^{16}\mbox{O}$. This quenching 
correction of the effective operator allowed to obtain 
a reasonable description of the
spectrum of $^{17}$F but failed in solving the
problem of 'halo' of discrete states for positive energies which was observed
in the elastic cross section and in the phase shifts \cite{ostatnia}. 
The whole problem
results from the non-hermitean corrections to the eigenvalues for positive
energies which generate the imaginary part and which are 
particularly large for pure single-particle
(or single-hole) configurations. For this reason, the incorrect 
internal configuration mixing in SM wavefunctions 
may lead to an unphysical enhancement
of the resonant-like correction from bound states into the scattering data
(e.g. the elastic phase-shifts). This aspect of the
continuum coupling we want to investigate using the ZBM interaction which
includes the most essential for the present studies configuration mixing in low
energy wavefunctions.

In Fig. 1 we show SMEC energies and widths  for
positive parity (l.h.s.\ of the plot) and negative parity (r.h.s.\ of the
plot) states of $^{17}\mbox{F}$. The calculations were performed using either the
density dependent interaction of Ref. \cite{schw84}, called DDSM0, 
or the similar interaction
with the overall strength reduced by a factor 0.67 , called DDSM1 
\cite{ostatnia}. (In both cases,  radius of the density formfactor has been
reduced as discussed above.) This latter interaction was used before in the 
SMEC calculations in $0p1s0d$ SM space \cite{ostatnia}. 
However, since the external mixing of 
wavefunctions in the SM space of ZBM is smaller than in the $0p1s0d$  SM space, 
therefore one may consider
bigger coupling strength of the residual interaction in SMEC-ZBM calculations.
For that reason, we compare results obtained with both
DDSM0 and DDSM1 interactions.
For the comparison, the experimental data and the SM input in
$Q$-space is shown in Fig. 1 separately for positive and negative
parity states as well. The agreement of the SM-ZBM calculations
with the experimental data is even better than obtained in the  
$0p1s0d$ SM space \cite{ostatnia}.
The agreement of SMEC calculations with experiment is encouraging for both
density dependent interactions but the widths of states are better reproduced by
SMEC-DDSM0. Only the coupling matrix
elements between the $J^{\pi}=0_{1}^{+}$ g.s.\ wavefunction of
$^{16}\mbox{O}$ and all considered states in $^{17}\mbox{F}$ are included.
Zero of the energy scale for $^{17}\mbox{F}$ is fixed by the experimental
position of $J^{\pi}=1/2_{1}^{+}$ ($E = - 105$ keV) 
with respect to the proton emission threshold. In Fig. 1 the theoretical
prediction for $3/2^{+}$ state is compared with the experimental $3/2_2^{+}$
state, because the s.p. orbit $0d_{3/2}$ is missing in SM space of ZBM.
For those s.p.\ wavefunctions which in a given many body state
are not modified by the selfconsistent
renormalization, we calculate radial formfactors using the common
reference s.p.\ potential $U^{(ref)}$ of the WS type 
which is adjusted to reproduce 
experimental binding energies of of $5/2_{1}^{+}$ and $1/2_{1}^{+}$ states 
for protons \cite{ostatnia}. This potential has the following 
parameters : the   radius $R_0 = 3.214\,$fm, the 
diffuseness $a = 0.58\,$fm, the spin-orbit strength $V_{SO} = 3.683\,$MeV and
the depth $V_0 = -52.46\,$MeV. The same potential without the Coulomb term is
also used to calculate radial formfactors for neutrons.
In the case when the self-consistent 
conditions in a $J^{\pi}$ state determine the
radial function of a s.p. wavefunction $\phi_{l,j}$, 
%for each residual interaction separately 
we readjust the depth of 'first guess' potential $U(r)$ so
that the energy of the s.p. state $\phi_{l,j}$ in the converged potential
$U^{(sc)}(r)$ corresponds to the energy of this s.p. state in $U^{(ref)}$. 
The remaining parameters : $R_0$, $a$, $V_{SO}$, of the initial potential 
are the same as in the reference potential $U^{(ref)}$. 
This readjustment of $V_0$ in the initial potentials $U(J^{\pi})$ for a
s.p. state $(l,j)$ guarantees that the binding of self-consistently determined
wavefunction $\phi_{l,j}$ is at the experimental value. 
Moreover, we have the same s.p.
energies and, consequently, the same asymptotic behaviours of wavefunctions 
for large $r$ for both residual density dependent interactions. 
This choice is essential for quantitative description of 
radiative capture cross-section.

$B(E2)$ transition matrix
element between $1/2_1^+$ and $5/2_1^+$ bound states is an interesting test
of the wavefunction. 
In SMEC with ZBM interaction and DDSM0 residual
coupling, the value for this
transition : $B(E2) = 79.61\,$e$^2$fm$^4$, which is obtained 
assuming the effective charges :
$e_p \equiv 1+\delta_p$, $e_n \equiv \delta_n$, with the polarization charge 
$\delta =\delta_p=\delta_n=0.2$, agrees with the 
 experimental value $B(E2)_{exp} = 64.92\,$e$^2$fm$^4$.
For the range of $\delta$ in between 0.1 and 0.3, which is
compatible with the theoretical estimates \cite{kirson}, the SM values with
Harmonic Oscillator wavefunctions are much smaller.
%$B(E2)_{SM} \in (18.7 - 26.7)$e$^2$fm$^4$, strongly underestimate the
%experimental value.  In the SM calculations we take 
%the oscillator length $b = 1.01 A^{1/6}\,$fm. 
This difference reflects a more realistic radial
dependence of the $1s_{1/2}$ s.p.\ orbit in $J^{\pi} = 1/2_1^{+}$ many body
state which is in part a consequence of the external mixing in SMEC
wavefunctions due to the coupling to the scattering continuum and provides the
halo structure of the $1s_{1/2}$ s.p.\ orbit. Similar effect can be generated
using the WS potential or P\"oschl-Teller-Ginocchio potential for the
appropriately chosen $1s_{1/2}$ orbit \cite{Gin84}. 
The rms radius for this orbit in SMEC
is : ${<r^2>}^{1/2} = 5.24\,$fm, as compared to : ${<r^2>}^{1/2} = 3.03\,$fm in
SM.

In Fig. 2, the calculated total astrophysical $S$-factor
as a function of the c.m.\ energy, as well as its values for the
$^{16}\mbox{O}(p,\gamma)^{17}\mbox{F}(J^{\pi}=1/2_1^{+})$ and
$^{16}\mbox{O}(p,\gamma)^{17}\mbox{F}(J^{\pi}=5/2_1^{+})$ branches 
are compared with the experimental data \cite{morlock}. Results shown in Fig. 2 have been
calculated with DDSM1 interaction. The total $S$-factor depends weakly on the
strength of the residual coupling and with DDSM0
interactions one obtains very similar results. The energy scale is adjusted to
reproduce the experimental position of 
$1/2_{1}^{+}$ state with respect to the proton emission threshold. Therefore,
the energy scale for excitation energy is the same as c.m. energy in the $p +
^{16}\mbox{O}$ system.
 The photon energy is then given by the difference of
c.m.\ energy of $[^{16}\mbox{O} + \mbox{p}]_{J_{i}}$~ system
and the experimental energy of the final state $J_f$ in $^{17}\mbox{F}$.
The dominant contribution to the total capture cross-section
for both $5/2_1^{+}$
and $1/2_1^{+}$ final states, is provided by $E1$ transitions from the
incoming $p$ wave to the bound $0d_{5/2}$ and $1s_{1/2}$ states. We took into
account all possible $E1$, $E2$, and $M1$ transitions from incoming $s$,
$p$, $d$, $f$, and $g$ waves but only $E1$ from incoming $p$ - waves
give important contributions. In the transition to the g.s., the $E1$
contribution from incoming $f_{7/2}$ wave is by
a factor $\sim$100 smaller than the contribution from $p_{3/2}$ wave 
at $E_{CM}\sim 0$. This contribution however increases with
the energy of incoming proton and becomes $\sim 0.3$ at 3.5~MeV.

The energy dependence
of $S$-factor as $E_{CM} \rightarrow 0$ can be fitted by a  second
order polynomial to calculated points obtained in the interval from 20 to
50~keV in steps of 1~keV. We have $S(0) = 9.32\times10^{-3}$~MeV$\cdot$b, 
and the logarithmic derivative is $S^{'}(0)/S(0) = -4.86$~MeV$^{-1}$. These
results have been obtained for DDSM1 residual interaction but are 
practically identical for the DDSM0 force.
The ratio of $E2$ and $E1$ contributions in the branch
$^{16}\mbox{O}(p,\gamma)^{17}\mbox{F}(J^{\pi}=1/2_1^{+})$ for DDSM1 interaction
is : ${\sigma}^{E2}/{\sigma}^{E1}=1.622\times10^{-4}$, $2.225\times10^{-4}$ and
$5.458\times10^{-4}$~ at 20, 100 and 500~keV, respectively. Also these ratios
do not depend on the strength of the residual coupling. On the contrary, the
ratio ${\sigma}^{M1}/{\sigma}^{E1}$ depends on the residual coupling and
equals 3.90$\times10^{-4}$ and 1.087$\times10^{-3}$ at $E_{CM} \rightarrow 0$
for DDSM1 and DDSM0 interactions respectively. At 500 keV, this ratio equals
6.11$\times10^{-5}$ and 1.81$\times10^{-4}$ for both interactions, 
respectively.

For the deexcitation to the g.s.\ $5/2_1^+$, the fit of calculated
$S$-factor for DDSM1 interaction as  $E_{CM} \rightarrow 0$ yields:
$S(0)=3\times10^{-4}$~MeV$\cdot$b and $S'(0)/S(0) =0.649$~MeV$^{-1}$.  
Almost identical results are obtained with the DDSM0 interaction. 
The ratio of $E2$ and $E1$ contributions for DDSM1 interaction 
at 20, 100 and 500~keV is:
${\sigma}^{E2}/{\sigma}^{E1}=1.336\times10^{-3}$, $1.11\times10^{-3}$ and 
$9.74\times10^{-4}$,  respectively. These ratios show some sensitivity
on the residual coupling. For DDSM0 one obtains : 
${\sigma}^{E2}/{\sigma}^{E1}=2.25\times10^{-3}$, $1.458\times10^{-3}$ and 
$1.043\times10^{-3}$ at 20, 100 and 500~keV,  respectively. 
Similarly as in the branch 
$^{16}\mbox{O}(p,\gamma)^{17}\mbox{F}(J^{\pi}=1/2_1^{+})$, the
ratio ${\sigma}^{M1}/{\sigma}^{E1}$ depends on the residual coupling and
equals 2$\times10^{-3}$ and 2.99$\times10^{-3}$ at $E_{CM} \rightarrow 0$
for DDSM1 and DDSM0 interactions respectively. 
In the total cross section, ratio of $E2$ and $E1$ contributions 
at 20, 100 and 500~keV equals :
${\sigma}^{E2}/{\sigma}^{E1}=2.03\times10^{-4}$,
$2.63\times10^{-4}$ and $5.85\times10^{-4}$ for 
DDSM1 interaction , and ${\sigma}^{E2}/{\sigma}^{E1}=2.34\times10^{-4}$,
$2.8\times10^{-4}$ and $5.92\times10^{-4}$ for 
DDSM0 interaction. 

Elastic phase shifts and elastic cross-sections for different
proton bombarding energies are shown in Figs. 3 and 4.
The elastic phase shifts data \cite{bluehab} are very 
well reproduced by the SMEC calculations for all considered 
partial waves including 
the $5/2^+$ for which a significant discrepancy with the data has been reported 
in SMEC calculations using the $p-sd$ interaction neglecting
higher order p-h correlations in the SM wavefunction for positive parity
states \cite{ostatnia}. In Fig. 3 we show the calculations for the DDSM1
residual interaction. As we have already mentioned, $0d_{3/2}$ s.p. orbit is
missing in the $Q$ subspace. On the other hand, the $3/2^+$ partial wave
contributes to the elastic cross-section. So we have decided to include this
resonance in the $P$ subspace, adjusting scattering potential for the $d_{3/2}$
proton wave to place it at the experimental position. The encouraging 
agreement of SMEC results with the data for the spectrum of
$^{17}\mbox{F}$, the proton capture cross-section 
$^{16}\mbox{O}(p,\gamma)^{17}\mbox{F}(J^{\pi}=5/2_1^{+})$ and the 
$5/2^+$ elastic phase shift is not accidental and shows importance
of the np-nh excitations across the 'closed core' $N=Z=8$, which are taken into
account in the present studies with the ZBM force. 

Calculated  elastic excitation functions at a laboratory angle of 166$^\circ $ 
in SMEC with DDSM0 (the solid line) and DDSM1 (the dashed line) 
are compared with the experimental data
\cite{salisbury} in Fig. 4. The agreement is very encouraging 
for both residual interactions in the almost entire energy range.
where the interference pattern depending sensitively 
on the precise values of the energy and width of resonance states is absent. 
Earlier SMEC calculations \cite{ostatnia}, using the simplified 
wavefunctions for the g.s. of $^{16}\mbox{O}$ and 
$^{17}\mbox{F}$, failed to
reproduce the experimental elastic cross-section in the low-energy domain below
the resonances. We have found that 
an unrealistic account for excitations from $0p$ to $1s0d$ shells
leads to a large resonant contribution from the g.s.\ $5/2^{+}_{1}$ 
wavefunction to the
phase shift in the partial wave $5/2^{+}$ and implies a strong decrease 
of the elastic excitation function at low energies. This effect is caused by
the virtual coupling of discrete and continuum states.
For energies below the proton emission threshold, coupling to
the continuum introduces hermitean modifications of $H_{QQ}$
which shift the energy of $5/2_1^+$ state with respect to its initial
position given by the SM but do not generate any width for this state.
For excitation energies above the proton threshold, the $Q - P$ coupling 
generates non-hermitean corrections which yield the imaginary part
of the eigenvalue of $H_{QQ}^{eff}$ and, hence, produce the resonant behavior.
Internal mixing of configurations in the SM wavefunctions tend 
to reduce this resonant behavior for positive energies  which, in general, 
is strongest for the states having little internal mixing, {\it i.e.} those
having a s.p. nature.  Actually, these strong resonant-like features
associated with certain bound states for positive energies 
(above the particle-emission
threshold) and the large shifts of the real part for certain eigenvalues 
at negative energies (below the particle-emission
threshold), have the same origin in
the interplay between external ({\it i.e.} via the
continuum coupling) and internal ({\it i.e.} within $Q$ space) 
configuration mixing in the SMEC wavefunction for this state.  
This example shows also that in the
SMEC approach, one may use different experimental observables
to fix those few parameters of the model such as the
overall strength of the residual $Q - P$ coupling or the radius and depth
of the initial average potential. Moreover, the information about
the amount of correlations in the low-lying coexisting 
states can be extracted not only
from the spectroscopic data but also from the elastic excitation function. 
For nuclei far from the stability
line where the amount of experimental data is strongly limited, this
feature of the model is very attractive. 

We feel that the evidence presented in
this work shows that the SM calculation extended to include the coupling to the
continuum of the scattering states can go a long way towards providing a
detailed explanation not only of the structure of $^{17}\mbox{F}$,  
$^{16}\mbox{O}$ and neighboring nuclei, but also the reaction data involving
one nucleon in the continuum. The corner-stone of this model is the effective
SM interaction providing a realistic 
internal mixing of configurations in the $Q$
subspace.  Several problems remain, such as 
the missing configurations and/or more realistic asymptotic decay
channels which show up in the decay width of resonances. In this latter case,
the extension of the SMEC is being investigated.

\vskip 1truecm

{\bf Acknowledgments}\\
We thank E. Caurier for his help 
in the early stage of development of SMEC
model. This work was partly supported by
KBN Grant No. 2 P03B 097 16 and the Grant No. 76044
of the French - Polish Cooperation.

%\newpage

\vfill
\newpage

\begin{figure}[t]
\centerline{\epsfig{figure=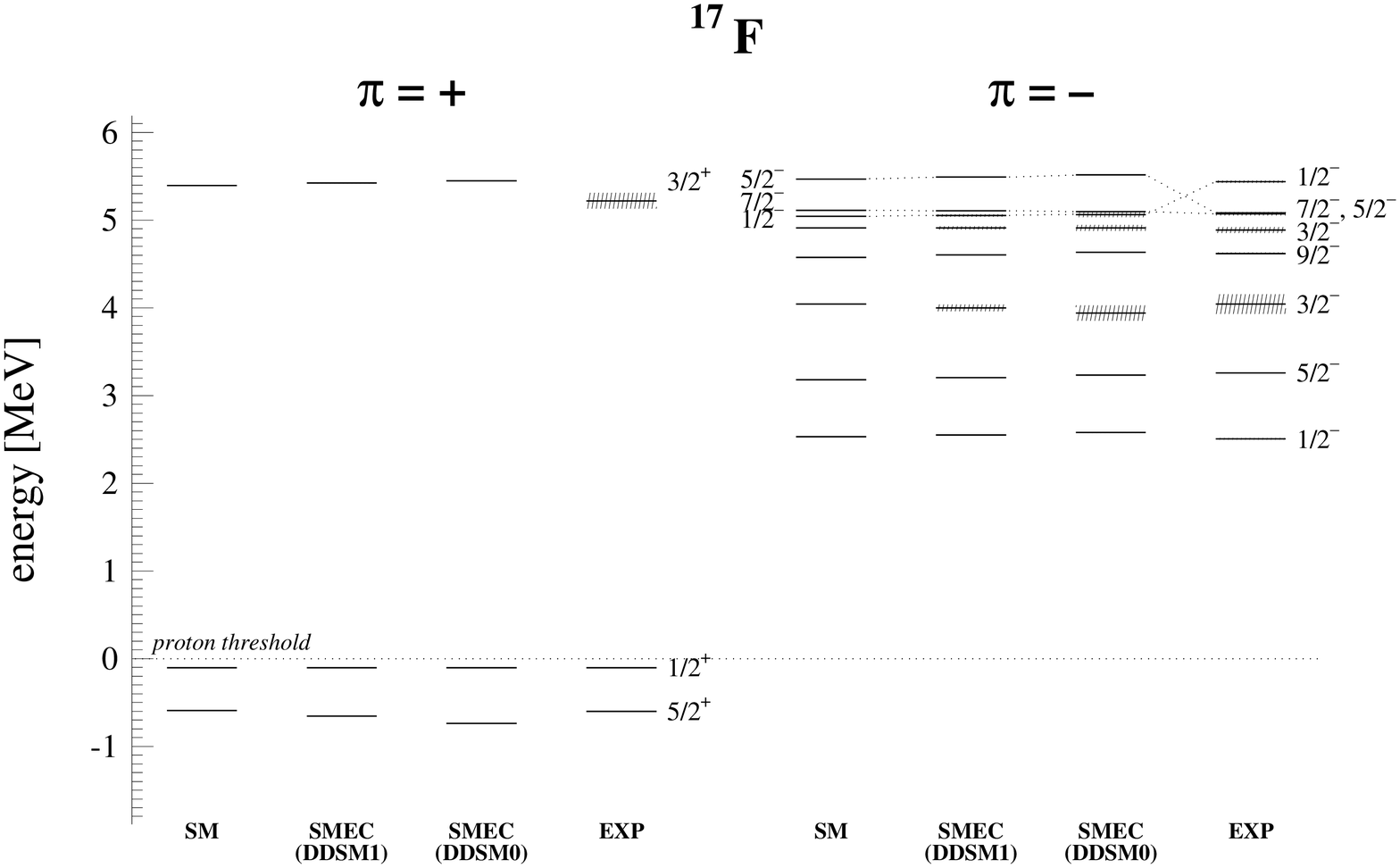,height=14cm}}
\vskip 1truecm                                                               
\caption{SM and SMEC energy spectra obtained with the ZBM effective SM
interaction
\protect\cite{zbm,zbm1} are compared with the experimental states of
$^{17}\mbox{F}$ nucleus. For the residual coupling between $Q$ and $P$
subspaces we use the density dependent DDSM0 \protect\cite{schw84} and DDSM1
\protect\cite{ostatnia} interactions. The proton threshold
energy is adjusted to reproduce position of the $1/2_{1}^{+}$ first excited
state. The shaded regions represent the width of resonance states. For the
details of the calculation see the description in the text.}
\label{figf17}
\end{figure}
%%%%%%%%%%
\begin{figure}[t]
\centerline{\epsfig{figure=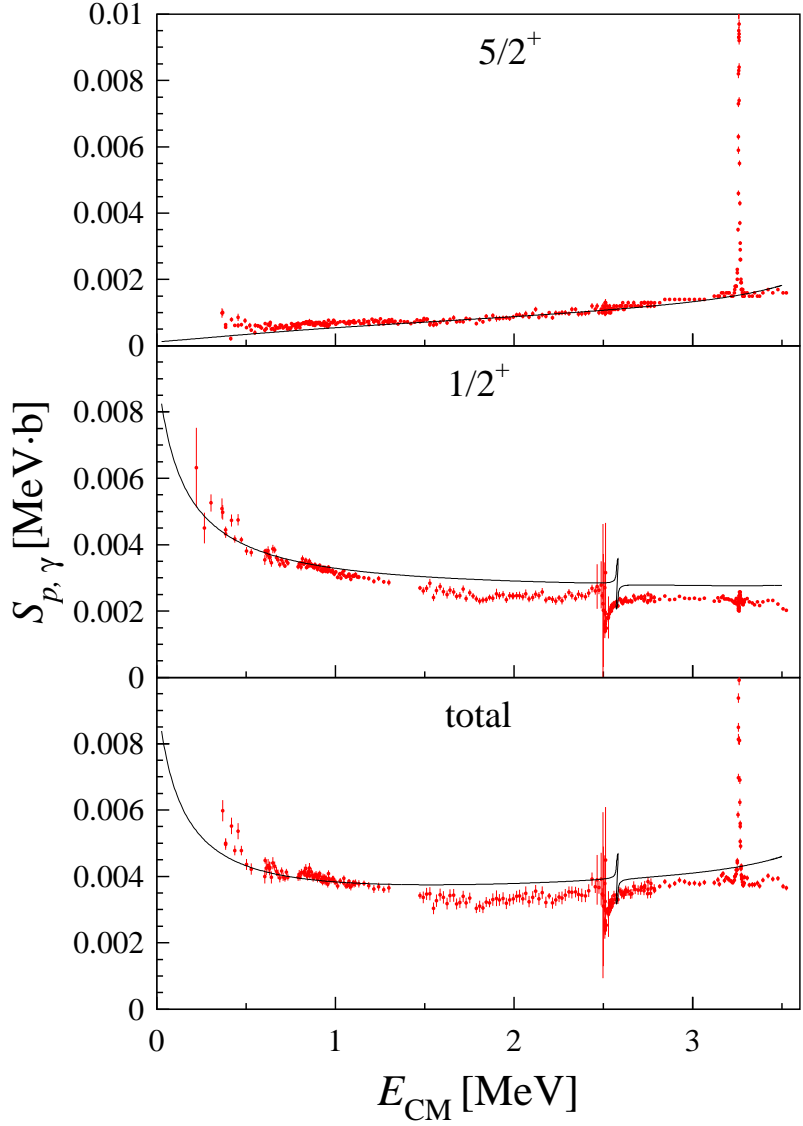,height=16cm}}
\vskip 1truecm
\caption{The astrophysical $S$-factor for the reactions
$^{16}\mbox{O}(p,\gamma)^{17}\mbox{F}(J^{\pi}=5/2_{1}^{+})$ and
$^{16}\mbox{O}(p,\gamma)^{17}\mbox{F}(J^{\pi}=1/2_{1}^{+})$ is plotted
as a function of the center of mass energy $E_{CM}$. For the residual coupling
between $Q$ and $P$ subspaces we use DDSM1 density dependent interaction
\protect\cite{ostatnia}. The experimental data are
from \protect\cite{morlock}. The contribution of $5/2_1^{-}$ resonance in  the
$^{16}\mbox{O}(p,\gamma)^{17}\mbox{F}(J^{\pi}=5/2_1^{+})$ branch is very
narrow in energy and has been omitted in the figure. The resonance is found at
5.29 MeV in the SMEC-DDSM1 calculations.}
\label{figa}
\end{figure}
%%%%%%%%%%
 \begin{figure}[t]
\centerline{\epsfig{figure=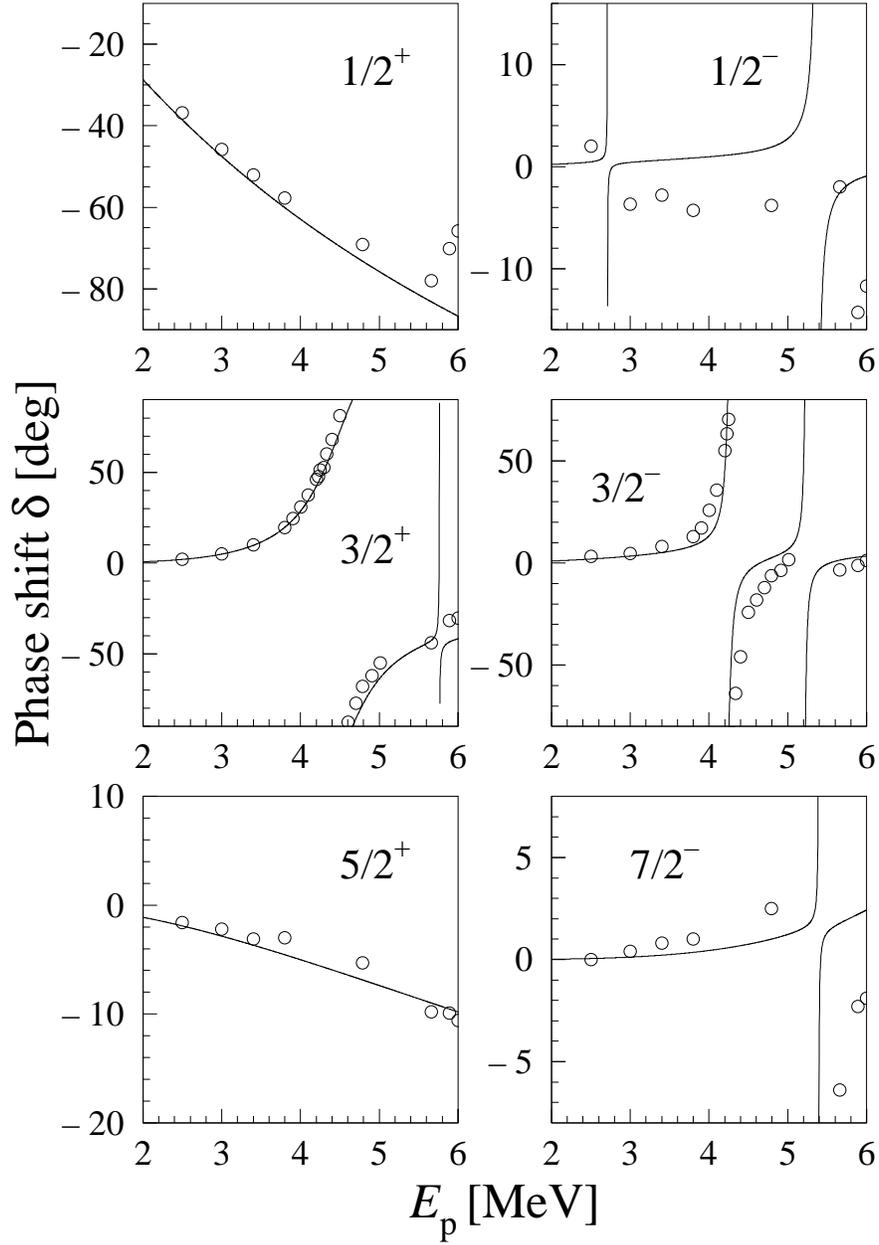,height=18cm}}
\vskip 1truecm
\caption{The phase-shifts for the $p + ^{16}\mbox{O}$ elastic scattering
as a function of the proton energy
$E_{p}$ for different partial waves.
The experimental data are from \protect\cite{bluehab}.
SMEC results have been obtained for the ZBM effective SM interaction
\protect\cite{zbm,zbm1} and the
density dependent residual interaction DDSM1 \protect\cite{ostatnia} 
(the solid line).}
\label{figc_1}
\end{figure}
%%%%%%%%%%
\begin{figure}[t]
\centerline{\epsfig{figure=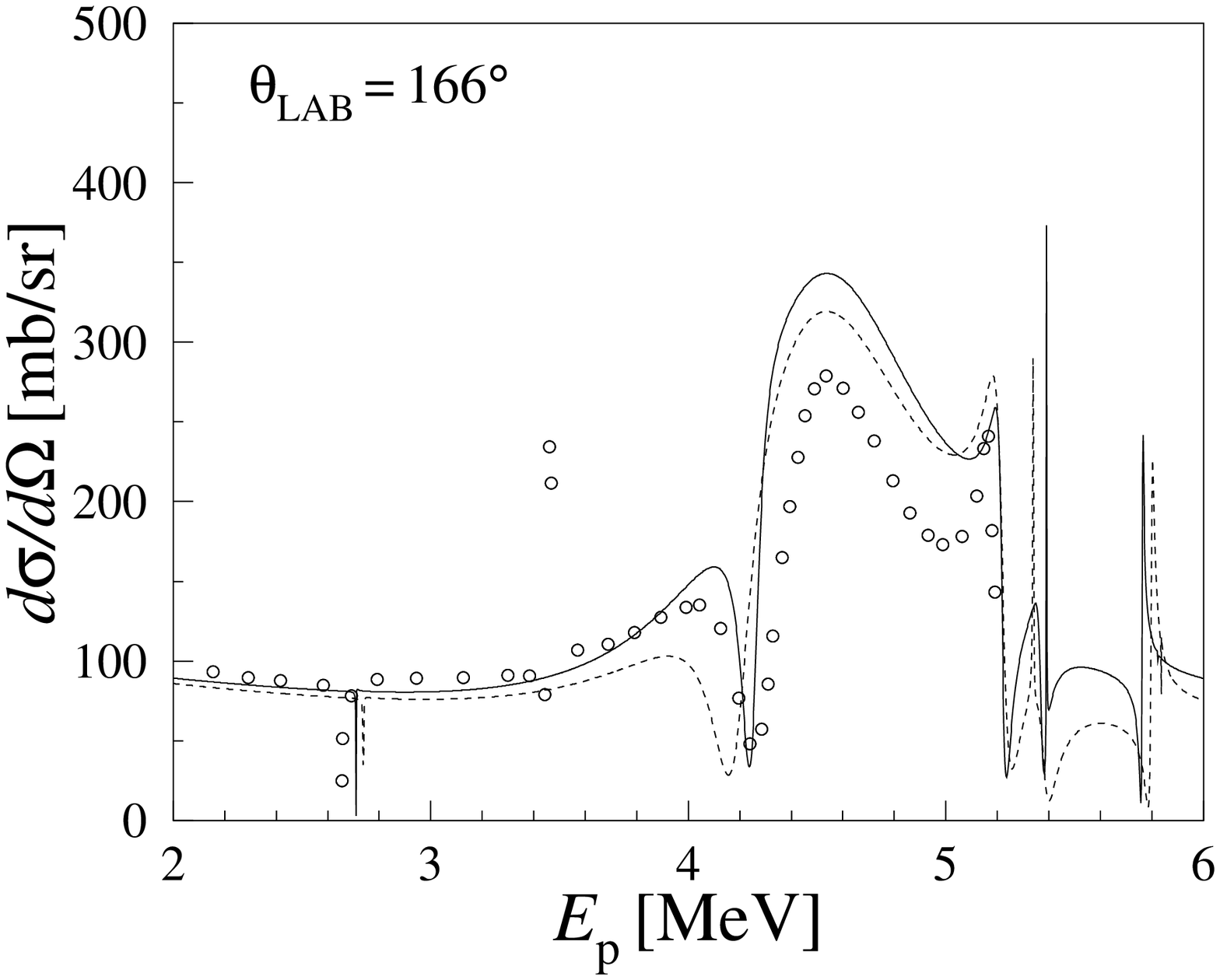,height=18cm}}
\vskip 1truecm
\caption{The elastic cross-section at a laboratory angle $\theta_{LAB} = 166
^\circ$ for the $p + {^{16}\mbox{O}}$ scattering as a function of proton
energy $E_p$. The SMEC calculations 
have been performed using either DDSM0 \protect\cite{schw84} 
(the dashed line) or DDSM1 \protect\cite{ostatnia} (the solid
line) residual interaction. Experimental
cross-sections are from Ref.~\protect\cite{salisbury}.}
\label{figb}
\end{figure}
%%%%%%%%%%


\begin{references}
\bibitem{comment}
The problem encountered by the ZBM interaction in describing $B(E1)$ 
transition strength is of no importance in the present analysis.
\end{references}
\end{document}